\documentstyle[epsfig,12pt]{article}

\columnwidth\textwidth

\newcommand{\comment}[1]{}

\newcommand{\bildchen}[3]{%                           % 
\begin{center}                                        %
\begin{flushleft}                                     %
\makebox{\Large $\displaystyle #2$}                   %
\end{flushleft}                                       %
\mbox{{\epsfig{figure=#1,width=10.cm,%                %
bbllx=1.8cm,bblly=9.2cm,bburx=20.cm,bbury=19.cm}}}   %
\end{center}                                          %
\begin{flushright}                                    %
   {\Large $\displaystyle #3$ \hspace*{5ex}}          %                  
\end{flushright}}                                     %
\begin{document}
\thispagestyle{empty}
\pagenumbering{arabic}
%-----------------------------
\everymath={\displaystyle}
\vspace*{-2mm}
\thispagestyle{empty}
\noindent
\hfill HUTP-96/A027\\
\mbox{}
\hfill hep-ph/9501286\\
\mbox{}
\hfill  Revised version: July 1996  \\
\begin{center}
  \begin{Large}
  \begin{bf} {\Large \sc 
{Radiative Corrections to $\pi_{l2}$ and $K_{l2}$ Decays}}
   \\
  \end{bf}
  \end{Large}
  \vspace{0.8cm}
   Markus Finkemeier \\[2mm]
   {\em Lyman Laboratory of Physics\\
        Harvard University\\
        Cambridge, MA 02138, USA}\\[5mm]
{\bf Abstract} % =====================================================
\end{center}
\begin{quotation}
\noindent
We reexamine radiative corrections to $\pi_{l2}$ and $K_{l2}$
decays.
We perform a matching calculation, including vector and axial
vector resonances as explicit degrees of freedom 
in the long distance part. 
By considering the dependence on the matching scale and on the hadronic
parameters, and by 
comparing with model independent estimates, we
scrutinize the model dependence of the results.
For the pseudoscalar meson decay constants, we extract the values
$f_\pi = (92.1 \pm 0.3)\,\mbox{MeV}$ and $f_K = (112.4
\pm 
0.9) \, \mbox{MeV}$. 
For the ratios $R_\pi$ and $R_K$ of the electronic
and muonic decay modes, we predict $R_\pi = (1.2354 \pm 0.0002) \cdot 
10^{-4}$ and $R_K = (2.472 \pm 0.001) \cdot 10^{-5}$.
\end{quotation}
\begin{center}
\end{center}
\newpage
%====================================================================
\vspace{5mm} {\bf 1.}
%====================================================================
Radiative corrections to $\pi_{l2}$ and $K_{l2}$ decays are interesting for
two separate reasons. On the one hand, measurements of the decay rates for
$\Gamma(\pi\to\mu\nu_\mu)$ and $\Gamma(K\to\mu\nu_\mu)$ are used to extract 
the decay constants $f_\pi$ and $f_K$, which are important input parameters
for chiral perturbation theory \cite{Gas84}. Therefore it is important to 
understand how radiative corrections affect these parameters \cite{Hol90}.
On the other hand, in the ratios $R_\pi = \Gamma(\pi\to e \nu_e)/
\Gamma(\pi \to \mu \nu_\mu)$ and $R_K$, 
strong interaction uncertainties cancel to a large degree. Therefore 
they can be predicted very precisely 
\cite{Ber58,Kin59,Ter73,Mar76,Gol77,Mar93}
and allow for
low energy precision tests of the standard model. 

The most recent discussion of radiative corrections to $\pi_{l2}$
decays can be found in \cite{Mar93}.
These authors seperate the radiative corrections into a long and a short
distance part, matched at a scale $m_\rho$.
In the long distance part, only pions are considered as active degrees of
freedom. 
QED corrections to the decay of a pointlike pion 
\cite{Kin59} give rise to the leading,  model independent
contribution.
Hadronic structure effects of the order 
${m_l^2}/{m_\rho^2} \ln (m_\rho^2 / m_l^2)$ 
are also model independent \cite{Ter73} and have been included.
For the remaining hadronic structure effects,
order-of-magnitude estimates are given.
Regarding the short distance part, the running of the effective semileptonic
four fermion interaction is used to evolve $G_F$ down from $m_Z$ to $m_\rho$.

This matching procedure is somewhat simplified, 
because in the long distance part
there are missing hadronic structure effects, which become important 
for energy scales approaching $m_\rho$, and in the short distance part,
the effective quark-antiquark-lepton-neutrino operator has been evolved 
down to the rather low scale $m_\rho$. 
We will use a different approach. We include vector and axial vector
resonances as explicit degrees of freedom in the long distance part,
which allows us to use a larger matching scale.
We show that this leads to a drastic reduction of the matching scale dependence
of the radiative correction.
The inclusion of the hadronic resonances, however, unavoidably 
introduces model dependence. Our main goal therefore will
be to study the size and the uncertainties of the model dependent 
contributions in detail, replacing the order-of-magnitude
estimates in \cite{Mar93}.

%====================================================================
\vspace{5mm} {\bf 2.}
%====================================================================
We seperate the loop integration over the Euklideanized momentum $k^2$
of the virtual photon into long distances
$k^2 = 0 \cdots \mu_{cut}^2$ and short distances
$k^2 = \mu_{cut}^2 \cdots m_Z^2$.
To calculate the long distance part,
we construct an effective model by starting with the low 
energy theorems of QCD and adding resonance degrees of freedom
along the lines of vector meson dominance.

The amplitude for the radiative decay $\pi\to l \nu_l \gamma$ 
consists of the model independent internal bremsstrahlung part (IB)
and the hadronic structure dependent (SD) part \cite{Bro64,Bij93}. 
The latter
is parametrized by two form factors $F_V(s)$ and $F_A(s)$.
$F_V(0)$ and $F_A(0)$ can be determined from chiral perturbation theory
\cite{Gas84,Bij93}
\begin{eqnarray}
   F_V^{(\pi)}(0) & = & \frac{m_\pi}{4 \sqrt{2} \pi^2 f_\pi}
\nonumber \\
%\nonumber \\
   F_A^{(\pi)}(0) & = & \frac{4 \sqrt{2} m_\pi}{f_\pi} (L_9 + L_{10}) 
\end{eqnarray}
We extrapolate from $s=0$ to $s \leq m_{\tau}^2$ by 
assuming dominance by low lying resonances with the correct quantum
numbers. In the case of $F_V$, we include small admixtures of the first two
higher radials, with $\lambda =0.136$ and $\mu=-0.051$ \cite{Dec93}.
Thus
\begin{eqnarray}
   F_V(s) & = & \frac{F_V(0)}{1 + \lambda + \mu} 
  \left[ \mbox{BW}_\rho(s) + \lambda \mbox{BW}_{\rho'}(s)
     + \mu \mbox{BW}_{\rho''}(s) \right]
\nonumber \\
   F_A(s) & = & F_A(0) \mbox{BW}_{a_1}(s)
\end{eqnarray}
$\mbox{BW}_X(t)$ denotes a Breit-Wigner propagator amplitude
with energy dependent widths  \cite{Kue90,Dec93b}. 
\begin{equation}
   \mbox{BW}_X(t) = \frac{m_X^2}{m_X^2 -t - i m_X \Gamma_X(t)}
\end{equation}
From this model for the amplitude $\pi\to l \nu_l \gamma$ we can derive
amplitudes for one-loop virtual corrections
by contracting the emitted photon back to the diagram in all possible ways,
using the same Feynman rules for the couplings of the photon again. 
This will be a good model for the virtual 
corrections for very small $k^2$ (where $k$ is the momentum flowing through
the photon), if the model for the radiative decay (where $k^2 = 0$)
was a realistic one in the first place.
To extrapolate these one-loop amplitudes from $k^2 \approx 0$ up to $k^2 
= \mu_{cut}^2$, we again use vector meson dominance. For the coupling
$\gamma \pi \pi$, we use the parameterization of \cite{Kue90} for the
electromagnetic form factor of the pion (which includes the $\rho$ and the 
$\rho'$). For the couplings of the photon to $\rho \pi$ and to $a_1 \pi$, we 
assume $\omega$ and $\rho$ dominance, respectively. 

In the case of the kaon $K \to l \nu_l (\gamma)$, we proceed very similarly.
$F_V^{(K)}(0)$ and $F_A^{(K)}(0)$ are obtained from chiral perturbation 
theory and extrapolated to higher $s$ assuming dominance by the
$K^*$ and the $K_1$, respectively. For the electromagnetic form factor
of the kaon we use a $1/2 : 1/6 : 1/3$ coherent superposition of
$\rho$, $\omega$ and $\Phi \sim (s \bar{s})$. 

For the short distance correction, arising from virtual photons with
$k^2 = \mu_{cut}^2 \dots m_Z^2$, we consider the one-loop running of the
effective four-fermion weak interaction 
$ [\overline{u}_{\nu_l} \gamma^\mu \gamma_- u_l]
  [\overline{u}_d \gamma_\mu \gamma_- u_u]$  from $m_Z$ down to
$\mu_{cut}$.
The 
leading $m_l$ and $\mu_{cut}$ dependence is given by \cite{Dec95}
\begin{equation} \label{eqn1}
   \left(\frac{\delta \Gamma}{\Gamma_0} \right)_{short\,\, dist.} \approx
   \frac{2 \alpha}{\pi} \frac{1}{m_l^2 - \mu_{cut}^2} 
   \left(m_l^2 \ln \frac{m_Z}{m_l} - \mu_{cut}^2 \ln \frac{m_Z}
   {\mu_{cut}} \right)
\end{equation}

%====================================================================
\vspace{5mm} {\bf 3.}
%====================================================================
%%%%%%%%%%%%%%%%%%%%%%%%%%%%%%%%% fig1
\begin{figure}
\caption{Radiative correction to $\Gamma(\pi\to\mu\nu_\mu)$, 
from our evaluation (solid) and from Ref.~\protect\cite{Mar93} (dashed)}
\label{fig1}
\bildchen{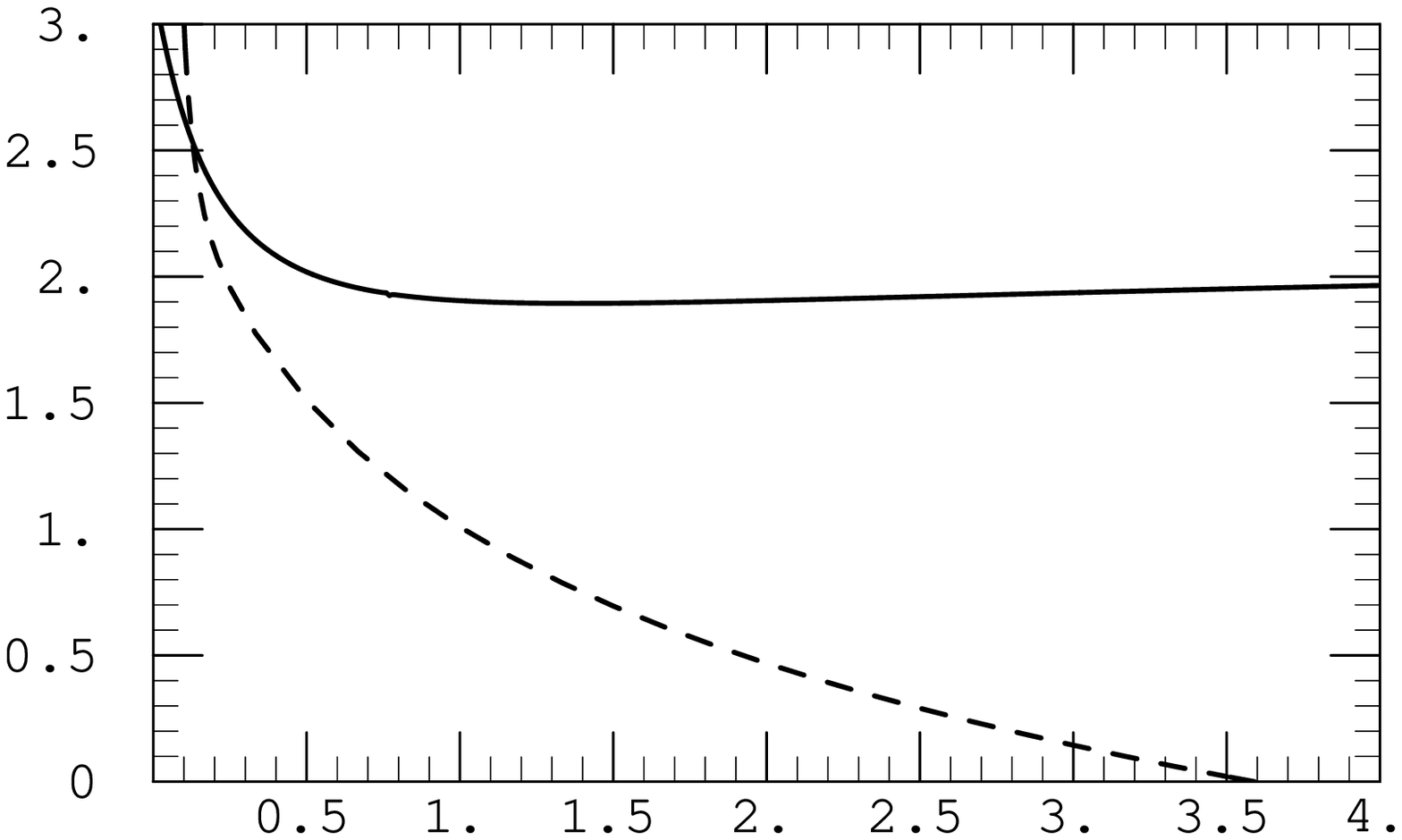}
{\frac{\delta \Gamma}{\Gamma_0}(\pi\to\mu\nu_\mu(\gamma)) [\%]}
{\mu_{cut}\,[\mbox{GeV}]}
\end{figure}
%%%%%%%%%%%%%%%%%%%%%%%%%%%%%%%%%%%%%%%%%%%%%%%%%%%%
In Fig.~\ref{fig1}, we present the numerical result for the radiative
correction ${\delta \Gamma}/\Gamma_0$ for the decay $\pi \to \mu \nu_\mu
(\gamma)$ in variation with the matching scale $\mu_{cut}$. 
(We include all photons in the radiative decay, without
a cut on the photon energy.)
We also compare to the corresponding result from \cite{Mar93}. It is seen
clearly that the inclusion of the meson resonances as explicit degrees
of freedom drastically reduces the matching scale dependence.
Our numerical result is stable for $\mu_{cut}$ from below
$0.5 \, \mbox{GeV}$ to well above $4 \, \mbox{GeV}$. 

From Fig.~\ref{fig1}, we obtain
\begin{eqnarray}
   \frac{\delta \Gamma}{\Gamma_0}(\pi\to\mu\nu_\mu(\gamma))
   & = & (1.88 \pm 0.04 \pm 0.08) \% + O(\alpha^2) + O(\alpha \alpha_s)
\end{eqnarray}
The central value $1.88\%$ has been obtained using $\mu_{cut}
= 1.5\, \mbox{GeV}$. We use a somewhat high central
value for $\mu_{cut}$, because
we have included the radial excitations ($\rho'$, $\rho''$) in the long 
distance part.
The first error quoted ($\pm 0.04 \%$) is the matching uncertainty, 
estimated by varying
$\mu_{cut}$ by a factor of two ($0.75 \dots 3 \, \mbox{GeV}$).
The second
error ($\pm 0.08\%$) estimate 
is the uncertainty from the hadronic parameters, obtained by
varying $F_{V,A}(0)$, the relative contributions of the higher resonances
and the resonance widths over reasonable ranges.

Leading higher order short distance corrections have been
estimated in \cite{Mar93}, which increase the short distance
correction by $0.10\%$.
There exist no estimates of $O(\alpha^2)$ corrections in 
the long distance part.

Therefore we will use 
\begin{eqnarray}
   \frac{\delta \Gamma}{\Gamma_0}(\pi\to\mu\nu_\mu(\gamma))
   & = & (2.0 \pm 0.2) \% 
\end{eqnarray}
to extract $f_\pi$. With $|V_{ud}| = 0.9744 \pm 0.0010$ \cite{RPP94}, we
obtain
\begin{equation}
\label{eqnfpi}
   f_\pi  
% = 93.015\, \mbox{MeV} \left[ 1 - \frac{(2.0 \pm 0.2)\%}{2} \right] 
  = (92.14 \pm 0.09 \pm 0.09) \, \mbox{MeV}
%  = (92.1 \pm 0.1) \, \mbox{MeV}
\end{equation}
where the first error, $\pm 0.09$, is due to $V_{ud}$, 
and the second one 
to the radiative corrections.

It should be emphasized that the definition of 
$f_\pi$ is not unambigous at $O(\alpha)$ \cite{Mar93}. 
By convention, one could absorb
part of the radiative correction in $f_\pi$. We define $f_\pi$ by 
factoring out {\em all} radiative corrections from $f_\pi$.
This convention is identical to the one used in \cite{Mar91,Mar93}, but
not to the one used in \cite{Hol90}.

Our result for $f_\pi$ has to be compared to the one in \cite{Mar93},
which is also quoted by the particle data group  \cite{RPP94}.
Transcribing their result to our convention and to 
$|V_{ud}| = 0.9744 \pm 0.0010$, their result reads
\begin{equation}
   f_\pi = (92.47 \pm 0.09 \pm 0.26) \, \mbox{MeV}
\end{equation}		
where the first error $\pm 0.09$ is due to $V_{ud}$, and the 
second error is estimated 
from the matching scale dependence. This is compatible
with our result (\ref{eqnfpi}).

In applications of $f_\pi$, one should use an error estimate 
which includes the full model dependence. Therefore we quote
\begin{equation}
   f_\pi = (92.1 \pm 0.3) \, \mbox{MeV}
\end{equation}
as our final result for $f_\pi$.

%%%%%%%%%%%%%%%%%%%%%%%%%%%%%%%%%%%%%% Fig.2
\begin{figure}
\caption{Radiative correction to the ratio $R_\pi$.
Solid: central values for the hadronic parameters.
Dashed and dotted: Reasonable variations of these parameters.}
\label{fig2}
\bildchen{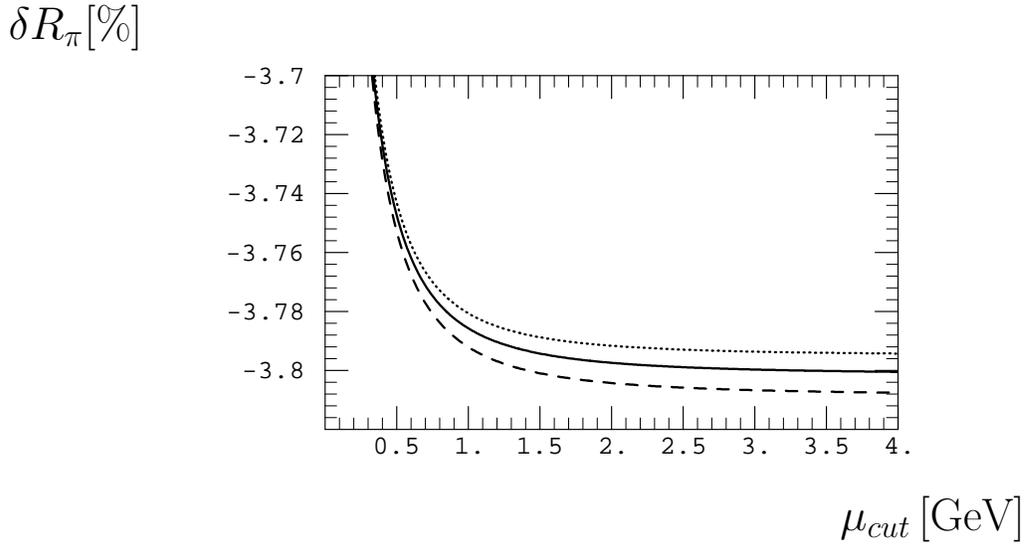}
{\delta R_\pi [\%]}
{\mu_{cut}\,[\mbox{GeV}]}
\end{figure}
%%%%%%%%%%%%%%%%%%%%%%%%%%%%%%%%%%%%%%
In Fig.~\ref{fig2}, we show our results for the radiative
correction to the ratio $R_\pi$. 
By convention, we have included all radiative decays
$\pi \to l \nu_l \gamma$ (IB + SD) in calculating $R_\pi$ (no cut
on the photon energy).
We obtain
\begin{equation} \label{eqnpi1}
  \delta R_\pi = - (3.793 \pm 0.019 \pm 0.007) \%
  + O(\alpha^2)
%  = - (3.793 \pm 0.020) \% + O(\alpha^2)
\end{equation}
where the first error ($0.019\%$) is the matching uncertainty,
estimated by varying $\mu_{cut}$ from $0.75$ up to $3 \, \mbox{GeV}$, and
the second error ($0.007\%$) arises from the uncertainties in the
hadronic parameters. 

To further study the model independence of the result, we have analyzed in
detail which scales contribute to the the loop integrals.
We find that the
contribution to the corrections to the decay rates themselves
remain sizable at large
$k^2$. However, the results for the electronic and muonic modes become
approximately equal for large $k^2$, and so the ratio $R_\pi$ is dominated
by very small scales. The total contribution within the range $\sqrt{k^2} = 
0.5 \cdots 3.0\, \mbox{GeV}$, where the theoretical uncertainties
are largest, is found to be $0.026 \%$ (we have added
absolute values to take care of cancellations).

We have also compared our result to the leading model independent
contributions \cite{Ter73} in detail. 
The hadronic structure dependent correction
can be separated into three parts, corresponding to three gauge invariant
sets of diagrams. 
Adding separately the absolute values of the differences of our full
results minus the model independent contributions for these three parts,
we obtain $0.011\%$.
In view of this, we consider the error estimate $\pm 0.020 \%$ in
(\ref{eqnpi1}) as reliable. 

We again have to consider higher order radiative corrections.
In \cite{Mar93}, the leading logarithms in
$\ln(m_\mu/m_e)$ have been summed up
to all orders in $\alpha$, increasing $R_\pi$ by
$5.5 \cdot 10^{-4}$.

And so our  prediction for $\delta R_\pi$  is
\begin{equation}
  \delta R_\pi = (-3.793 \pm 0.020 + 0.055 \pm 0.01) \%
  = (-3.74 \pm 0.03) \%
\end{equation}
In the sum, the first number is the central value 
and the second number 
the uncertainty of the $O(\alpha)$ correction.
The third number  is the leading higher order correction and
$\pm 0.01\%$ is our estimate of the next-to-leading correction.

For the ratio $R_\pi$, this implies
\begin{equation}
   R_\pi = R_\pi^{(0)} \Big( 1 + \delta R_\pi \Big)
%   = 1.2834 \cdot 10^{-4} \times \Big(1 - 0.0374 \pm 0.0002 \Big)
   = (1.2354 \pm 0.0002) \cdot 10^{-4}
\end{equation}
This agrees with the prediction $R_\pi = (1.2352 \pm 0.0005)\cdot 10^{-4}$
in \cite{Mar93} within their error estimate.

From a similar analysis for kaon decays, we obtain
$\frac{\delta \Gamma}{\Gamma_0}(K\to\mu\nu_\mu(\gamma)) = (1.3 \pm 0.2)\%$,
which using $|V_{us}| = 0.2205 \pm 0.0018$ \cite{RPP94}
results in
\begin{equation}
   f_K = (112.4 \pm 0.9 \pm 0.1) \mbox{MeV}
\end{equation}
The first error is due to $V_{us}$ and the second one to 
the radiative correction.

In calculating $R_K$, we include only the
(soft) internal bremsstrahlung  (IB) part of $K\to e \nu_e \gamma$ 
and exclude the (hard) structure dependent (SD) part.
Experimental results can be corrected to comply with this convention
using the theoretical results for the differential distributions
for the IB and the SD radiation \cite{Bro64,Bij93}.
We obtain
\begin{eqnarray}
   R_K
 & = &
   \frac{\Gamma(K \to e \nu_e (\gamma))}
   {\Gamma(K \to \mu \nu_\mu (\gamma))} 
\nonumber \\
%\nonumber \\
 & = &
 R_K^{(0)} \Big( 1 + \delta R_K \Big)
   =  2.569 \cdot 10^{-5} \times \Big(1 - 0.0378 \pm 0.0004 \Big)
\nonumber \\
%\nonumber \\
 &  = &
(2.472 \pm 0.001) \cdot 10^{-5}
\end{eqnarray}

%====================================================================
\vspace{5mm} {\bf 4.}
%====================================================================
We have calculated $f_\pi$, $f_K$, $R_\pi$ and $R_K$ in an
improved matching calculation, which includes vector and axial
vector resonances as explicit degrees of freedom.
The central values we quote
include small model dependent contributions, but their error
bars are based on the full size of these model dependent
contributions, and in this sense our final predictions
can be considered as model independent.

\section*{Acknowledgement}
The author would like to thank J. Bijnens, J. Gasser and 
J. Stern for helpful comments, and the members of the Theorectical
Physics Group at Harvard University for their kind hospitality.
This work is supported in part by the National Science Foundation
(Grant \#PHY-9218167) and by the Deutsche Forschungsgemeinschaft. 

%================================================================

\end{document}